\documentclass[preprint,12pt]{elsarticle}

\usepackage{amssymb}
\usepackage{amsthm}
\usepackage{caption}
\usepackage{geometry}
\usepackage[section]{placeins} 

\begin{document}

\begin{frontmatter}



\title{The collective motion of self-propelled particles affected by the spatial-dependent noise}


\author{Jia-xin Qian}

\author{Yan-qing Lu\corref{cor1}}
\ead{yqlu@nju.edu.cn}

\cortext[cor1]{Corresponding author}

\affiliation{organization={National Laboratory of Solid State Microstructures, Collaborative Innovation Center of Advanced Microstructures, College of Engineering and 		Applied Sciences},
	addressline={Nanjing University}, 
	city={Nanjing},
	postcode={210093}, 
	country={China}}

\begin{abstract}
We study the collective motion of self-propelled particles affected by the spatial-dependent noise based on the Vicsek rules.
Only the particles inside the special region will affected by noise.
The consideration of the spatial-dependent noise is closer to reality because of the complexity of the environment.
Interestingly, we find that there exists an optimal amplitude of noise to adjust the average motional direction of the system.
Particular orientation of the noisy region makes the motional direction of the system parallel to the orientation of the noisy region.
The adjustment of the motional direction of the system also depends on the shape, the proportion and the spatial distribution of the noisy region.
Our findings may inspire the capture of the key features of collective motion underlying various phenomena.
\end{abstract}




\end{frontmatter}

\section{Introduction}
Collective behavior of active agents is universally observed in rich scale of the systems from macroscopic to microscopic including the crowds of human\cite{RN47}, the schools of fish\cite{RN49} and the colonies of bateria\cite{RN48} etc\cite{RN45}.
Because of the complexity of the environment, most of the time, the behavior of agents will inevitably encounter noise.
Noise is present in the study of many domains\cite{RN40, RN31, RN42, RN11, RN22, RN9, RN19, RN26, RN27} and it plays an important role on the dynamics of the system including the system of active matter\cite{RN32, RN25, RN24}.
Studying how different kinds of noise affect the collective motion and what the noise can do in the system of active matter is useful in exploring the basic principles of collective motion, taking advantage of the features of noise and avoiding the unexpected disturbance caused by noise.

One of the typical models to studying collective motion, named Vicsek model, is proposed by Vicsek et.al in 1995\cite{RN46}.
In Vicsek model, all of the self-propelled particles follow the rule of velocity alignment which considers both the average velocity of the neighbor of particles and normal-distributed random noise to update the velocity of the particles\cite{RN46, RN21}.
Following Vicsek et.al, many researchers show interest in studying the collective motion with different kinds of noise including cross-correlated noise\cite{RN18, RN36, RN37, RN43}, non-Gaussian noise\cite{RN29}, colored noise\cite{RN28, RN38}, Telegrahic-like noise\cite{RN30} and hybrid noise\cite{RN44} etc\cite{RN39}.
And many exotic phenomena resulting from noise are found.
Noise can induce the transition of the motional state of the system\cite{RN10, RN13, RN20} and lead to symmetry breaking\cite{RN33}.
Driving the particles\cite{RN12, RN35} and maximizes collective motion in heterogeneous media are achieved by noise\cite{RN17}.
Noise also affects the criticality\cite{RN16}, synchronization\cite{RN15} and the diversity of collective motion in Vicsek model\cite{RN23}.

Although the studies on collective motion with noise make great progress in the recent decades, just a few studies consider the effect of the spatial feature of noise on collective motion\cite{RN34}.
Studying the collective motion with noise distributed nonuniform in space is an important step to futher understanding the key factor underlying the various phenomena of collective motion.
Therefore, we pay attention to the role of the spatial-dependent noise in the collective motion based on the Vicsek rules.
The update of the velocity of the particles will be affected by noise just when the particles are inside the special region.
By investigating the effect of the amplitude of noise, the orientation, the shape, the proportion and the spatial distribution of the noisy region on the average motional direction of the system, we find that the proper spatial-dependet noise can perfectly make the motional direction of the system parallel to the orientation of the noisy region.

\section{Model and method}
We consider $\mathit{N}$ particles move in the square cell with periodic boundary condition. 
The linear size of the square cell is $\mathit{L}$ and the particles are regarded as points.
Initially, particles are randomly distributed in the cell.
The position of all the particles update simultaneously at each time step $\mathit{\Delta t}$ following
\begin{equation}
	\mathbf{x_{i}}(t+\Delta t) = \mathbf{x_{i}}(t) + \mathbf{v_{i}}(t) \Delta t
\end{equation}
where $\mathbf{x_{i}}(t)$ and $\mathbf{v_{i}(t)}$ denote the position and velocity of the particles $\mathit{i}$ ($\mathit{i}$ takes $\mathit{1}$ to $\mathit{N}$) at time $\mathit{t}$ respectively.
The initial direction of velocity of the particles are distributed in $\mathit{[-\pi,\pi]}$ randomly and normally.
The magnitude of velocity of each particle is $\mathit{v}$.
Based on the rules of velocity alignment in Vicsek model, the update of the direction of velocity is as follow
\begin{equation}
	\theta_{i}(t+\Delta t) = Arg[\sum_{j\in \mathcal{N}_{i}(t)} e^{i\theta_{j}(t)}] + \Delta \theta(t)
\end{equation}
$\theta_{i}(t)$ denotes the direction of the velocity of particle $\mathit{i}$ at time $\mathit{t}$.
$\mathit{\mathcal{N}_{i}(t)}$ is the set of neighbors of the particle $\mathit{i}$ which means $\mathit{\mathcal{N}_{i}(t)} = \{ j : |\mathbf{x_{i}} - \mathbf{x_{j}}| \leqslant r \}$ and $\mathit{r}$ is the interaction radius of each particle.
$\mathit{\Delta \theta}$ denotes the noise.

Considering the effect of complex environment on motion, we take the spatial-dependent noise into account, which means
\begin{equation}
	\Delta \theta(t) = \eta(\mathbf{x_{i}}(t)) \xi_{i}(t)
\end{equation}
where $\mathit{\xi_{i}(t)}$ is a random number normally distributed in $\mathit{[-1/2, 1/2]}$.
When particle $\mathit{i}$ is in the rectangular region in green(as shown in Fig.1), $\mathit{\eta(\mathbf{x_{i}}(t)) = \eta_{n}}$ where $\mathit{\eta_{n}}$ is the amplitude of noise.
Otherwise, $\mathit{\eta(\mathbf{x_{i}}(t)) = 0}$.
\begin{figure}[!htb]
	\centerline{\includegraphics[width = 0.4\linewidth]{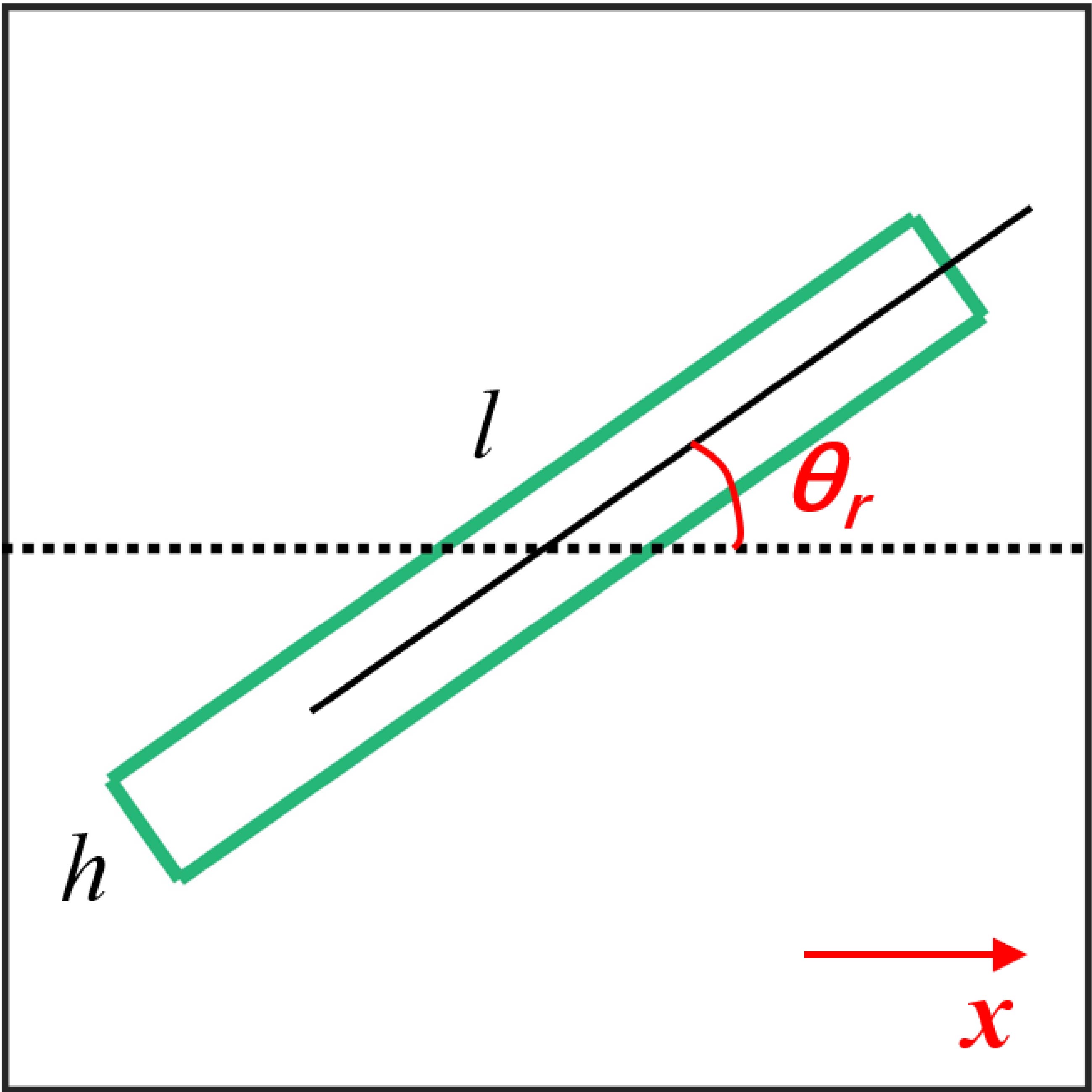}}
	\caption{The schematic diagram of our model with spatial-dependent noise. $\mathit{l}$ and $\mathit{h}$ denote the length and width of the noisy region respectively. And the orientation of the noisy region is denoted by the angle $\mathit{\theta_{r}}$.}
\end{figure}
As Fig. 1 shows, the orientation of the rectangular region is described by the angle $\mathit{\theta_{r}}$ between the long axis of the region and the positive direction of the X-axis which is shown by the red arrow.
The length and width of the rectangular region are $\mathit{l}$ and $\mathit{h}$ respectively.

To characterize the feature of the collective behavior, normalized average velocity is introduced as the order parameter, which is
\begin{equation}
	\phi = \frac{1}{Nv} \left | \sum_{i=1}^{N} \mathbf{v_{i}} \right |
\end{equation}

\section{Result and discussion}
In the simulation, $\mathit{\Delta t = 1}$, $\mathit{L = 10}$, $\mathit{v = 0.04}$.
When $\mathit{\eta_{n} = 0}$, the system will finally arrive to a state that all the particles move in the same direction as shown in Fig. 2(a).
\begin{figure}[!htb]
	\centerline{\includegraphics[width = 0.9\linewidth]{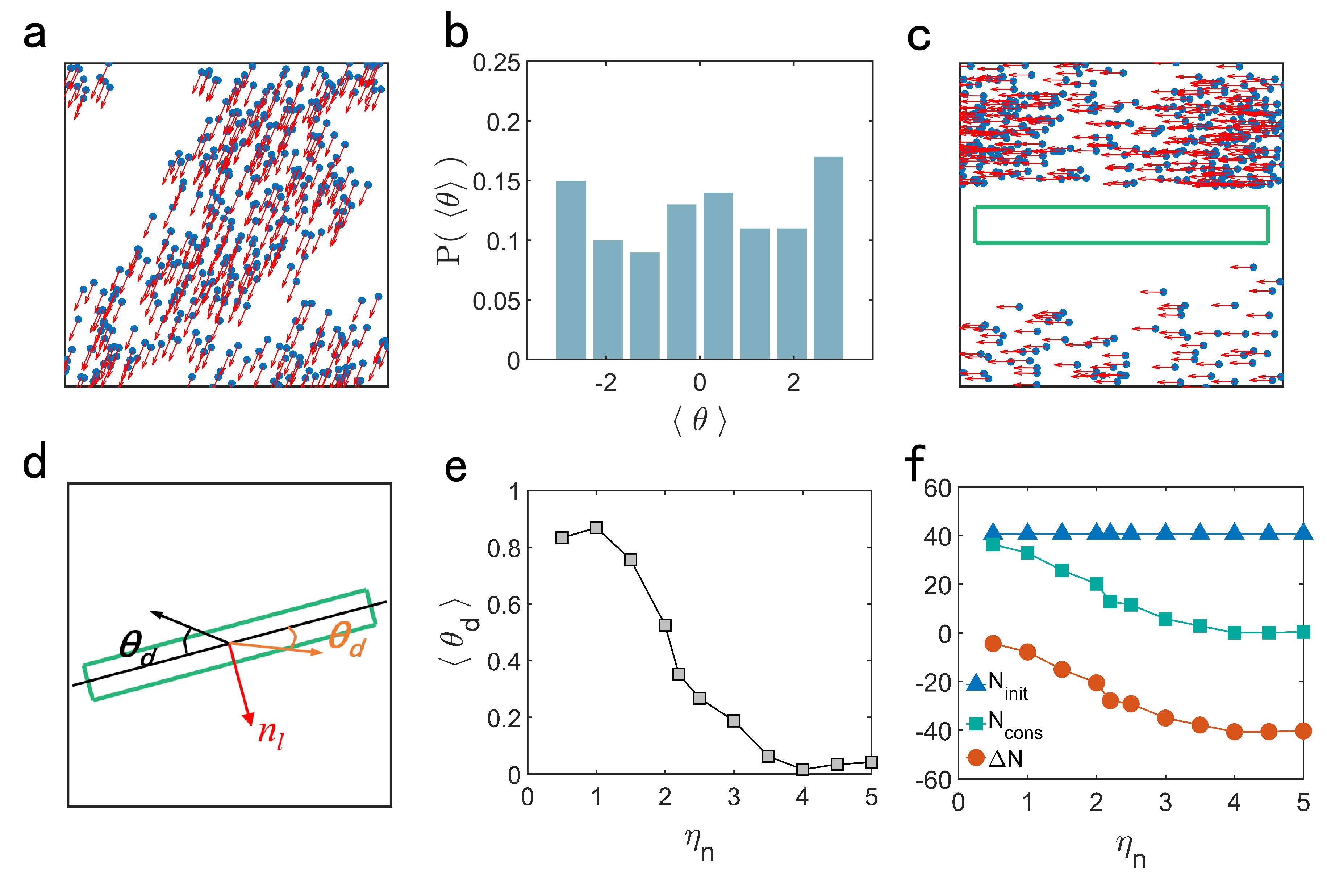}}
	\caption{The effect of the amplitude of the spatial-dependent noise on the adjustment of motional direction of the system.
		(a) the snapshot of all the particles move in almost the same direction without noise($\mathit{\eta_{n}}$);
		(b) the probability distribution of the average motional direction $\mathit{\langle \theta \rangle}$ without noise;
		(c) the snapshot of the system moves with the effect of spatial-dependent noise (the rectangular region in green is the noisy region);
		(d) the schematic diagram of the deviation $\mathit{\theta_{d}}$ between the average motional direction of the system $\mathit{\langle \theta \rangle}$ and the orientation of the noisy region;
		(e) the deviation $\mathit{\theta_{d}}$ as a function of the amplitude of noise $\mathit{\eta_{n}}$;
		(f) the number of the particles in the noisy region at the beginning of the simulation $\mathit{N_{init}}$, reaching motional consensus $\mathit{N_{cons}}$ and the different $\mathit{\Delta N = N_{cons} - N_{init}}$ as a function of $\mathit{\eta_{n}}$.}
\end{figure}
Fig .2(b) shows the probability distribution of the average direction of $\mathit{100}$ realizations when $\mathit{\eta_{n} = 0}$, which implies the motional direction of the system in each realization is random.
In the present of the spatial-dependent noise, $\mathit{\eta_{n} \neq 0}$.
We observe that the motional direction is parallel to the orientation of the rectangular region as shown in Fig. 2(c).

In order to know why the spatial-dependent noise can adjust the direction of the system, we first investegate the effect of the amplitude of noise $\mathit{\eta_{n}}$ on the motion of the system.
To measure the deviation between the average motional direction and the orientation of the rectangular region with noise, we consider the angle $\mathit{\theta_{d}}$.
As shown in Fig. 2(d), the vector $\mathbf{n}_{l}$ perpendicular to the length of the rectangular region is define as its normal vector.
If the angle between the average motional direction $\mathit{\langle \theta \rangle}$ and the normal vector $\mathbf{n}_{l}$ is larger than $\mathit{0.5\pi}$, the deviation is $\mathit{\theta_{d} = \langle \theta \rangle - 0.5\pi}$ as the black angle $\theta_{d}$ shown in Fig. 2(d).
Otherwise, $\mathit{\theta_{d} = 0.5\pi - \langle \theta \rangle}$ as the orange angle $\theta_{d}$ shown in Fig. 2(d).

As Fig. 2(e) shows, the average deviation of the $\mathit{100}$ realizations decreases as the amplitude of noise increases.
When $\mathit{\eta_{n} = 4}$, the average deviation is minimum, which means the best control of the direction of moltion.

To understand how the spatial-dependent noise can adjust the motional consensus of the system, we study the number of particles in the noisy region in different stages of the simulations.
As shown in Fig. 2(f), with the increasing of $\mathit{\eta_{n}}$, the number of particles when the system reaches motional consensus outside the noisy region $\mathit{N_{cons}}$ decrease.
And $\mathit{N_{cons}}$ is minimum when $\mathit{\eta_{n} = 4}$.
So do the different $\mathit{\Delta N = N_{cons} - N_{init}}$.
That means all the particles will move outside the noisy region when the amplitude of spatial-dependent noise is optimal.
It leads to the adjustment of the average motional direction which is parallel to the orientation of the noisy region.
When the amplitude of the noise outside the noisy region is small, it will have the similar phenomena and rules.

Beside the amplitude of the noise, the orientation of the noisy region is also important to the adjustment of the average motional direction.
We respectively investegate the average deviation of the motional direction $\mathit{\langle \theta_{d} \rangle}$ and the probability when $\mathit{\langle \theta_{d} \rangle} \leqslant 0.03$.
As shown in Fig. 3(a) and (b), the average deviation of the motional direction is smaller than $\mathit{0.4}$ with various orientation of noisy region.
\begin{figure}[!htb]
	\centerline{\includegraphics[width = 0.65\linewidth]{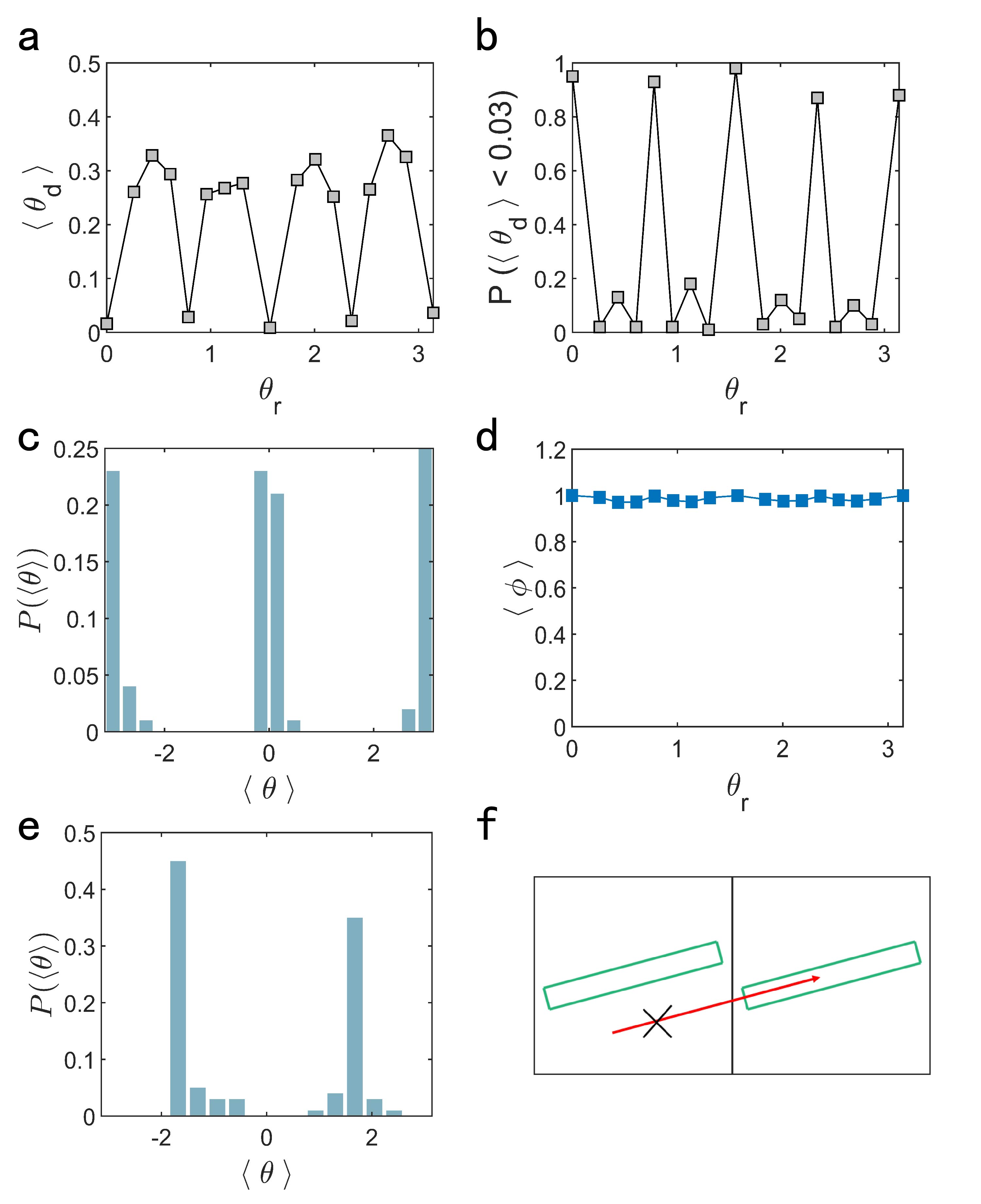}}
	\caption{The effect of the orientation of the noisy region on the adjustment of the motional direction of the system.
		(a) the deviation $\mathit{\theta_{d}}$ as a function of the orientation of the noisy region $\mathit{\theta_{r}}$;
		(b) the probability that $\mathit{\theta_{d}}$ is smaller than 0.03 as a function of $\mathit{\theta_{r}}$;
		The probability distribution of the average motional direction $\mathit{\langle \theta \rangle}$ when $\mathit{\theta_{r} = 0.083\pi}$(c) and $\mathit{\theta_{r} = 0.583\pi}$(e).
		(d) the average order parameter $\mathit{\langle \phi \rangle}$ as a function of $\mathit{\theta_{r}}$;
		(f) the schematic diagram of the average motional direction is not parallel to the orientation of the noisy region because off the avoidence to move into the noisy region.}
\end{figure}
And the system moves parallel to the orientation of noisy region when $\mathit{\theta{r} = 0, 0.25\pi, 0.5\pi, 0.75\pi, \pi}$.

Although the average order parameter shown in Fig. 3(d) is large enough to denote the motion consensus of the system.
The average direction of the system is not always parallel to the orientation of the noisy region.
For example, when $\mathit{\theta_{r} = 0.083\pi}$ and $\mathit{\theta_{r} = 0.583\pi}$, the average motional direction of the system is horizontal and vertical respectively, as the probability distribution of the average motional direction shown in Fig. 3(c) and Fig. 3(e) respectively.
The system will not prefer to move parallel to the orientation of the noisy region except $\mathit{\theta{r} = 0, 0.25\pi, 0.5\pi, 0.75\pi, \pi}$.
Because of the periodic boundary condition, it will make particles move inside the noisy region as Fig. 3(f) shows.
And the system will be adjusted to move in the direction that the particles will not move inside the noisy region.

The shape of the noisy region also has an impact on the motional direction of the system.
Keeping the same proportion of the noisy region as $\mathit{p = 0.1}$, we change the length $\mathit{l}$ of the rectangular noisy region to investegate the effect of shape on the motional direction.
The length of the rectangular noisy region in Fig. 4(a),(c) and (e) are $\mathit{l = 8.0}$, $\mathit{l = 5.0}$ and $\mathit{l = 3.0}$ respectively.
\begin{figure}[!htb]
	\centerline{\includegraphics[width = 0.9\linewidth]{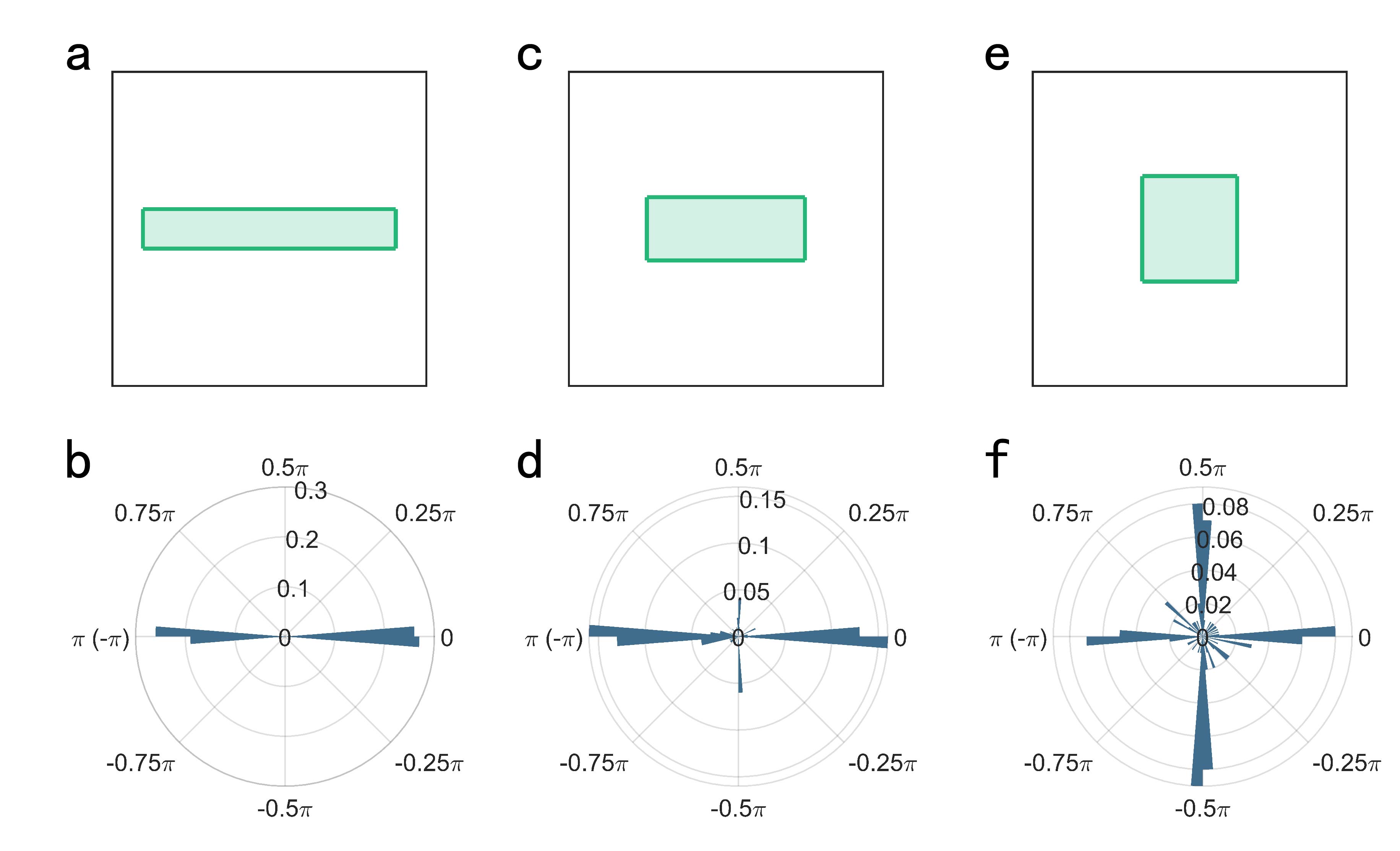}}
	\caption{The effect of the shape of the noisy region on the adjustment of the motional direction of the system.
		The schematic diagram of the rectangular noisy region with length $\mathit{l = 8.0}$(a), $\mathit{l = 5.0}$(c) and $\mathit{l = 3.0}$(e).
		The polardiagram of the probability distribution of the average motional direction $\mathit{\langle \theta \rangle}$ when the length of the noisy region is $\mathit{l = 8.0}$(b), $\mathit{l = 5.0}$(d) and $\mathit{l = 3.0}$(f).}
\end{figure}
With the decreasing of the length, the noisy region gradually changes from a rectangle to a square.
The motional direction of the system is from horizontal to both vertical and horizontal.
When $\mathit{l = 5.0}$, most of the motional direction is horizontal and a few motional directions is vertical because of the available space for the system to move vertically outside the noisy region.

Considering the proportion of the noisy region affect the degree of adjustment of the motional direction, we study the effect of proportion of the noisy region by investeagting the probability of realizations when the ratio of the particles moving horizontally $\mathit{s}$ is larger than $\mathit{95}$ percent.
As shown in Fig. 5(a), the noisy region can not adjust the motional direction of the system at all when the proportion of the noisy region is larger than $\mathit{0.7}$.
\begin{figure}[!htb]
	\centerline{\includegraphics[width = 0.7\linewidth]{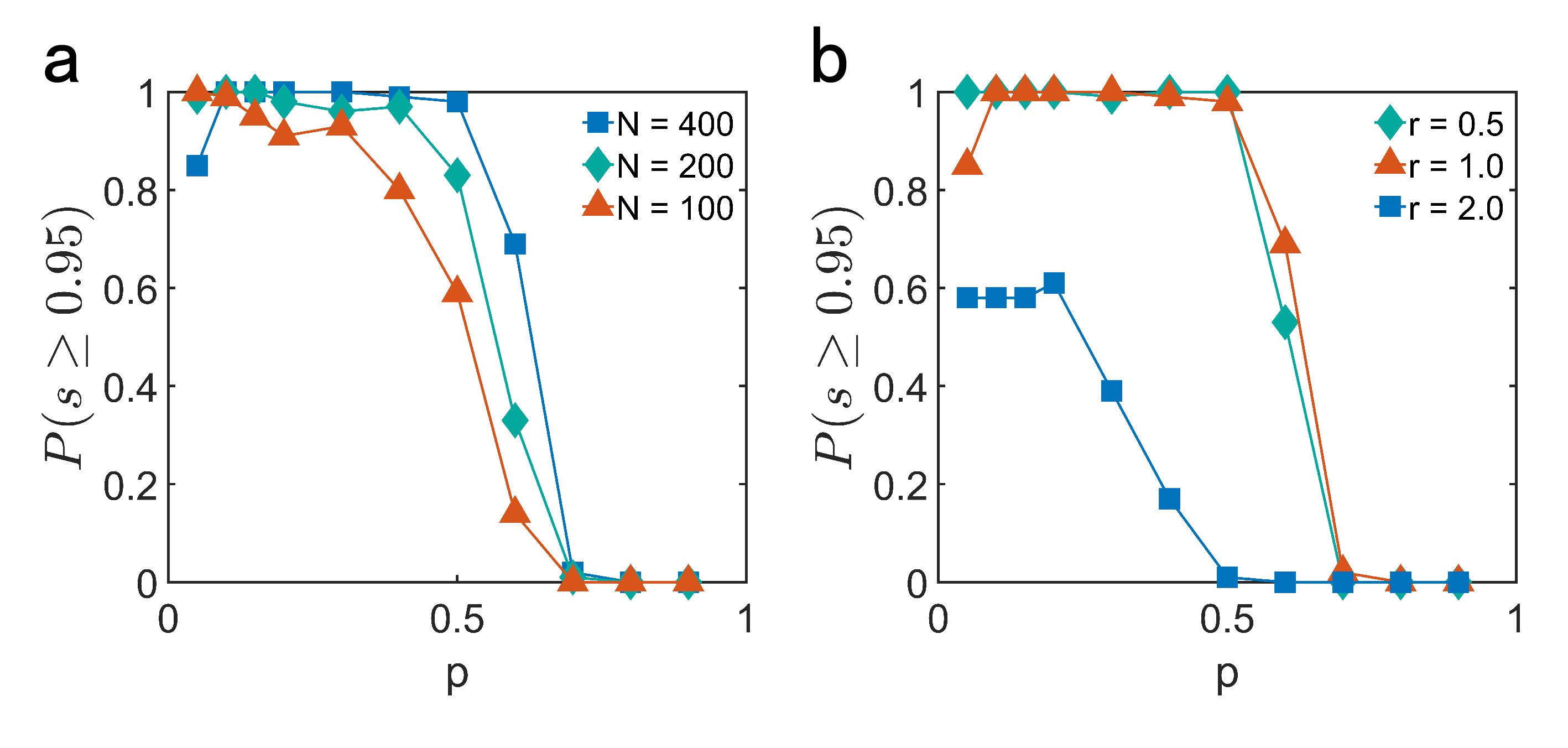}}
	\caption{The effect of the proportion of the noisy region on the adjustment of the motional direction of the system.
		The probability that the ratio of particles which move parallel to the orientation of the noisy region is larger than $\mathit{95}$ percent $\mathit{P(s \geqslant 0.95)}$ as a function of the proportion of the noisy region $\mathit{p}$ with different total number of particles(a) and different interaction radius(b).}
\end{figure}
And the degree of adjustment to the motional direction decreases as $\mathit{p}$ increase.
Large interaction radius, such as $\mathit{r = 2.0}$, weakens the adjustment on motional direction which makes the adjustment unstable as shown in Fig. 5(b).

As for the different spatial distribution of the noisy region, we set two equal-proportion noisy regions in the square cell and study the adjustment of motional direction of the system.
Both one noisy region and two noisy regions, with the same proportion, can adjust the motional direction of the system as shown in Fig. 6(c) and (d), but the degree of adjustment is different.
\begin{figure}[!htb]
	\centerline{\includegraphics[width = 0.7\linewidth]{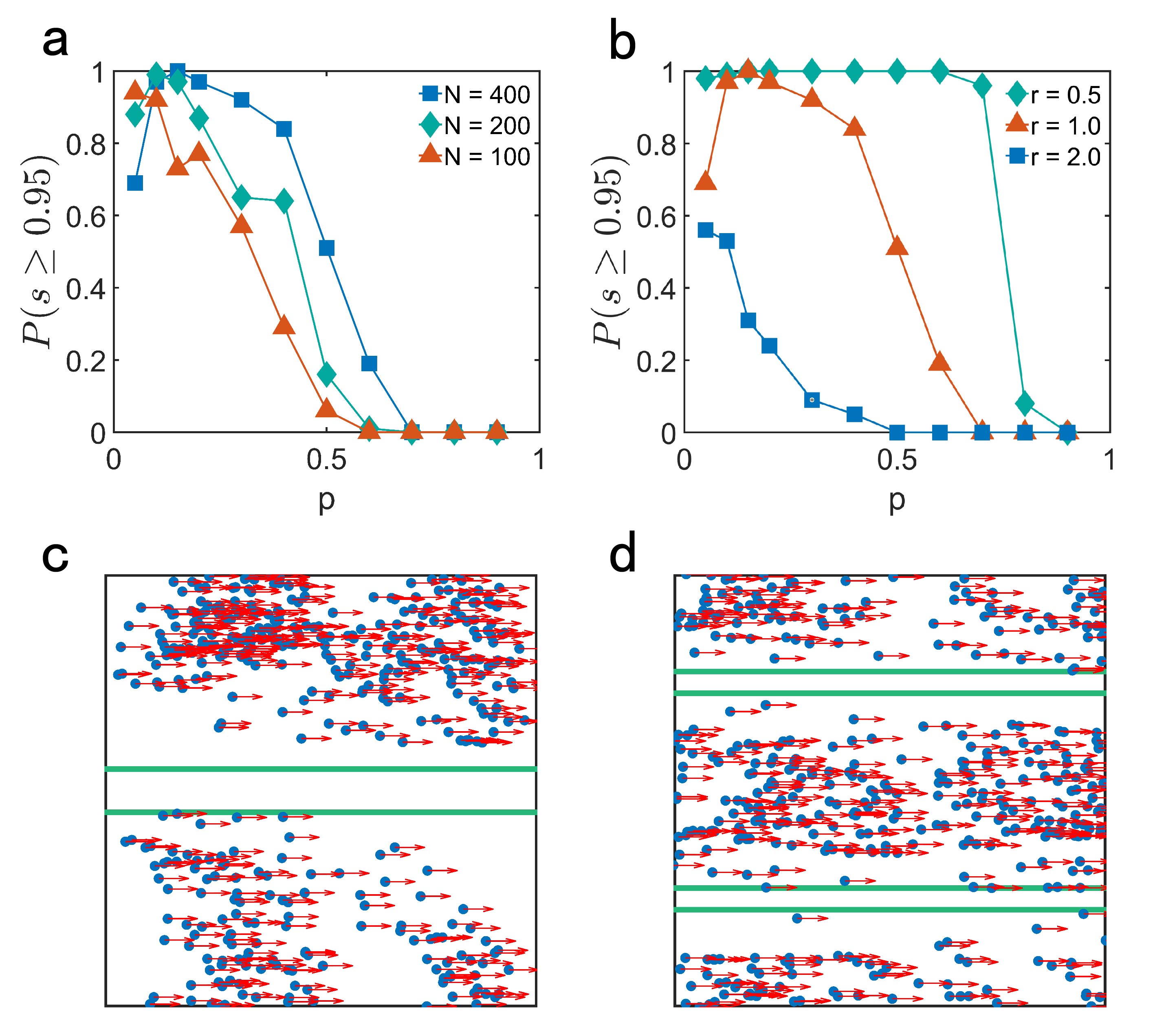}}
	\caption{The effect of total proportion of the two noisy regiona on the adjustment of the motional direction of the system.
		The degree of adjustment $\mathit{P(s \geqslant 0.95)}$ as a function of $\mathit{p}$ with different total number of particles(a) and different interaction radius(b).
		The snapshots of the particles move in the square with one noisy region(c) and two noisy regions(d). Total proportion of the noisy region in (c) and (d) is $\mathit{0.1}$.}
\end{figure}

As Fig. 6(a) shows, the two noisy region can hardly adjust the motional direction of the system when $\mathit{p}$ is larger than $\mathit{0.6}$.
This value is smaller than that when it is just one noisy region, which implies the weaker ability to adjust the motional direction of the system compared to one noisy region with the same proportion.
The difference of the ability to adjust the motional direction for different interaction radius is larger than that with one noisy region as shown in Fig. 6(b).

\section{Conclusion}
In conclusion, we study the collective motion of self-propelled particles affected by spatial-dependent noise based on vicsek rules.
The motion of the particles inside the rectangular region of the square cell will be affected by noise.
While other particles will move without noise.

Our investigation reveals that spatial-dependent noise enables to adjust the average motional direction of the system rather than moving in the random direction.
And the amplitude of noise, orientation, shape, proportion and spatial-distribution of noisy region have different impacts on the adjustment of the average motional direction.

It exsits the optimal amplitude of noise to achieve that all the particles move parallel to the orientation of the noisy region.
When the orientation of the rectangular noisy region are $\mathit{\theta_{r} = 0, 0.25\pi, 0.5\pi, 0.75\pi, \pi}$, the average motional direction of the system is peraller to the orientation of the noisy region, which means the spatial-dependent noise controls the motional direction of the system.
The motional direction of the system from just horizontal to both vertical and horizontal with the change in the shape of the noisy regoin from rectangle to square.

The degree of the adjustment of the motional direction decreases as the proportion of the noisy region increases both for one noisy region and two noisy regions.
Large interaction radius weaken the degree of adjustment of the motional direction.
The difference of adjustment between different $\mathit{r}$ is larger with two noisy regions although the total proportion of the noisy region is equal.

Our research may encourage further studies on the collective behavior in various and complex environments.

\section{Acknowledgments}
This work was supported by the Natural Science Foundation of Jiangsu Province, Major Project (Grants No. BK20212004).
The authors gratefully appreciate Prof. Bing-Xiang Li and Dr. Ling-Ling Ma for their constructive discussions and valuable help.




\bibliographystyle{elsarticle-num} 
\bibliography{ref_3}





\end{document}